\documentclass[12pt]{elsart}
\usepackage{epsfig}
\usepackage{amssymb}

\begin{document}
\begin{frontmatter}

\title{Quark pair simulation near threshold : Reweighting and hadronization}
\author{M. Boonekamp \thanksref{email}}
\thanks[email]{e-mail address: Maarten.Boonekamp@cern.ch}
\address{CE-Saclay, F-91191 Gif-sur-Yvette Cedex, France}

\begin{abstract}

Methods are presented allowing a realistic, experiment-based simulation
of low-mass quark pairs coupling to a virtual photon. Differential
cross-section reweighting factors allow to reproduce the resonant
structure observed near quark pair production thresholds, and an
algorithm is proposed that allows hadronization of the light quarks in
the very low mass region, where existing theory-based models are expected
to break down. The corresponding routines can be embedded in two- and
four-fermion generators for $e^{+}e^{-}$ machines.

\end{abstract}

\end{frontmatter}

\section{Introduction}

Several physics subjects of interest require a correct description of
low mass quark-antiquark systems. In $e^{+}e^{-} \rightarrow \mathrm{q\bar{q}}$
processes, initial state radiation can bring back the mass of the
outgoing quark pair near its production threshold, giving rise to
topologies with an energetic photon and a collimated hadronic
system. Such events can for example be used to extract a measurement
of the hadronic cross-section at the effective center-of-mass energy
(and the ratio R of hadronic to pointlike muonic cross-sections)
\cite{R_ISR}, or represent a background to searches for physics beyond
the standard model, such as radiative decays of intermediate vector bosons.

In neutral-current $e^{+}e^{-} \rightarrow \mathrm{q\bar{q} f\bar{f}}$
processes, low-mass quark pairs give again access to hadronized
virtual photons. Such events are interesting in their own right (they
have been studied by the LEP experiments at LEP1 \cite{ADLO_LEP1}, and
the corresponding cross-section is measured by the Delphi experiment
at LEP2 \cite{Delphi_LEP2}), but can also represent a background to
new particle searches, of which low mass Higgs boson searches are an
example.

Current two- and four-fermion generators are not adapted to these kinds of
studies. A first point is that the string model \cite{string}, upon which
the {\tt PYTHIA} program \cite{pythia} relies to simulate the
hadronization of parton systems, describes the data well at high energy,
but is expected to break down below about 2 GeV. Secondly, computations
of amplitudes generally make use of state-of-the-art knowledge of electroweak
corrections and include a first order QCD term, but ignore
non-perturbative corrections to vertices involving a light
$\mathrm{q\bar{q}}$ pair. In other words, diquark systems are generated
with distorted mass distributions in the low mass region, and are
hadronized unrealistically.

Low mass quark-antiquark pairs appearing in $e^{+}e^{-}$ annihilations
most often involve a coupling to a virtual photon. This suggests
using experimental data on $\mathrm{e^{+}e^{-} \rightarrow hadrons}$ a
low energy, namely the R-ratio and measurements of exclusive final states cross-sections, to solve
the problem. The mass spectrum and hadronization aspects are described
in turn in the next two sections. Although it is not the main subject
of this report, some results for two- and four-fermion processes are
summarized in the third section. The code location is given in
the last section.


\section{Quark-pair production near thresholds}

The task of implementing non-perturbative QCD corrections to
$\mathrm{q\bar{q}}$ differential cross-sections was first attempted in
the framework of the four-fermion generator {\tt FERMISV}
\cite{fermisv,janot}. The present method is similar and improves on a
few shortcomings : the integration of the $\mathrm{q\bar{q}}$ are
spectra are improved using proper Monte-Carlo techniques ; most
importantly, recent data in the charm resonance region are used in
order to obtain a good description of this mass range.

\subsection{Reweighting of the $\gamma^{*} \mathrm{q\bar{q}}$ vertex}

With QCD turned off, the cross-section for quark pair production through 
a virtual photon can be written (in a symbolic way):
\begin{eqnarray}
d\sigma^0_{hadrons} &=& d\sigma^0_{\mathrm{u\bar{u}}}
                     +  d\sigma^0_{\mathrm{d\bar{d}}}
                     +  d\sigma^0_{\mathrm{s\bar{s}}}
                     +  d\sigma^0_{\mathrm{c\bar{c}}}
                     +  d\sigma^0_{\mathrm{b\bar{b}}} \nonumber\\
&=& d\sigma_{\mu\mu} \times (R^0_u+R^0_d+R^0_s+R^0_c+R^0_b) \nonumber\\
&=& d\sigma_{\mu\mu} \times R^0, \nonumber
\end{eqnarray}
\noindent where the $R^0_q$ are just charge, colour and threshold
factors, already taken into account in generators, and $R^0$ is the
total parton level R-ratio. We now consider QCD effects, and write
similarly: 

\begin{eqnarray}
d\sigma_{hadrons}
&=& d\sigma_{\mu\mu} \times R \nonumber\\
&=& d\sigma_{\mu\mu} \times (R_u+R_d+R_s+R_c+R_b) \nonumber\\
&=& d\sigma^0_{\mathrm{u\bar{u}}} \bar{R}_u
  + d\sigma^0_{\mathrm{d\bar{d}}} \bar{R}_d
  + d\sigma^0_{\mathrm{s\bar{s}}} \bar{R}_s
  + d\sigma^0_{\mathrm{c\bar{c}}} \bar{R}_c
  + d\sigma^0_{\mathrm{b\bar{b}}} \bar{R}_b, \nonumber
\end{eqnarray}

\noindent with $\bar{R}_q \equiv R_q/R^0_q$ .The above lines express that for a
given quark flavour q, a QCD-improved description can be implemented by
multiplying the parton level matrix elements by a factor $\bar{R}_q$,
representing the contribution of quark flavour q to the experimental 
R-ratio, with charge, colour and threshold factors divided 
out\,\footnote{Of course, the splitting of R into contributions from
different flavours is only a computational convenience, and the whole
procedure makes sense only after summing all
contributions.}. Cross-sections of processes containing a
$\mathrm{q\bar{q}}$ pair will thus be computed as 

\begin{eqnarray}
d\sigma = \bar{R}_q(m_{\mathrm{q\bar{q}}}) \times |{\mathcal M}|^{2} \times dPh, \nonumber
\end{eqnarray}

\noindent with $|{\mathcal M}|^{2}$ the parton level matrix elements, and $\bar{R}_q$
evaluated at the $\mathrm{q\bar{q}}$ mass.

The parametrizaton of R includes the contributions from all
$J^{PC}=1^{--}$ resonances up to the $\Upsilon(11020)$, with parameters
taken from \cite{pdg}. Intuitively (and as in \cite{janot}), the
$\rho$'s and $\omega$'s are attributed to the u- and d-quarks ($ie$,
entered in the definitions of $\bar{R}_u$ and $\bar{R}_d$); the
$\phi$'s, the $\psi$'s and the $\Upsilon$'s are attributed to s-, c-
and b-quarks (included in $\bar{R}_s$,$\bar{R}_c$,$\bar{R}_b$) respectively.

The continuum is parametrized by simple functions satisfying partonic
boundary conditions, with parameters adjusted in the threshold regions
so that the available data \cite{Rdata} are correctly reproduced. At
present, one such function is defined for u- and d-quarks, one for
s-quarks, and one for c-quarks. Since the currently available data do
not allow a detailed representation of the Upsilon resonance region,
the continuum from b-quarks is represented by a step function at the
B-meson threshold. 

Finally, the continuum parametrizations include an overall
second order perturbative QCD correction factor. The running of
the strong coupling constant is computed at second order, and
$\Lambda_{QCD}$ is set so as to reproduce the world-averaged
$\alpha_S(m_{Z}^{2})$ \cite{pdg}. The implementation used is taken
from \cite{pythia}.

\begin{figure}
  \begin{center}
    \epsfig{figure=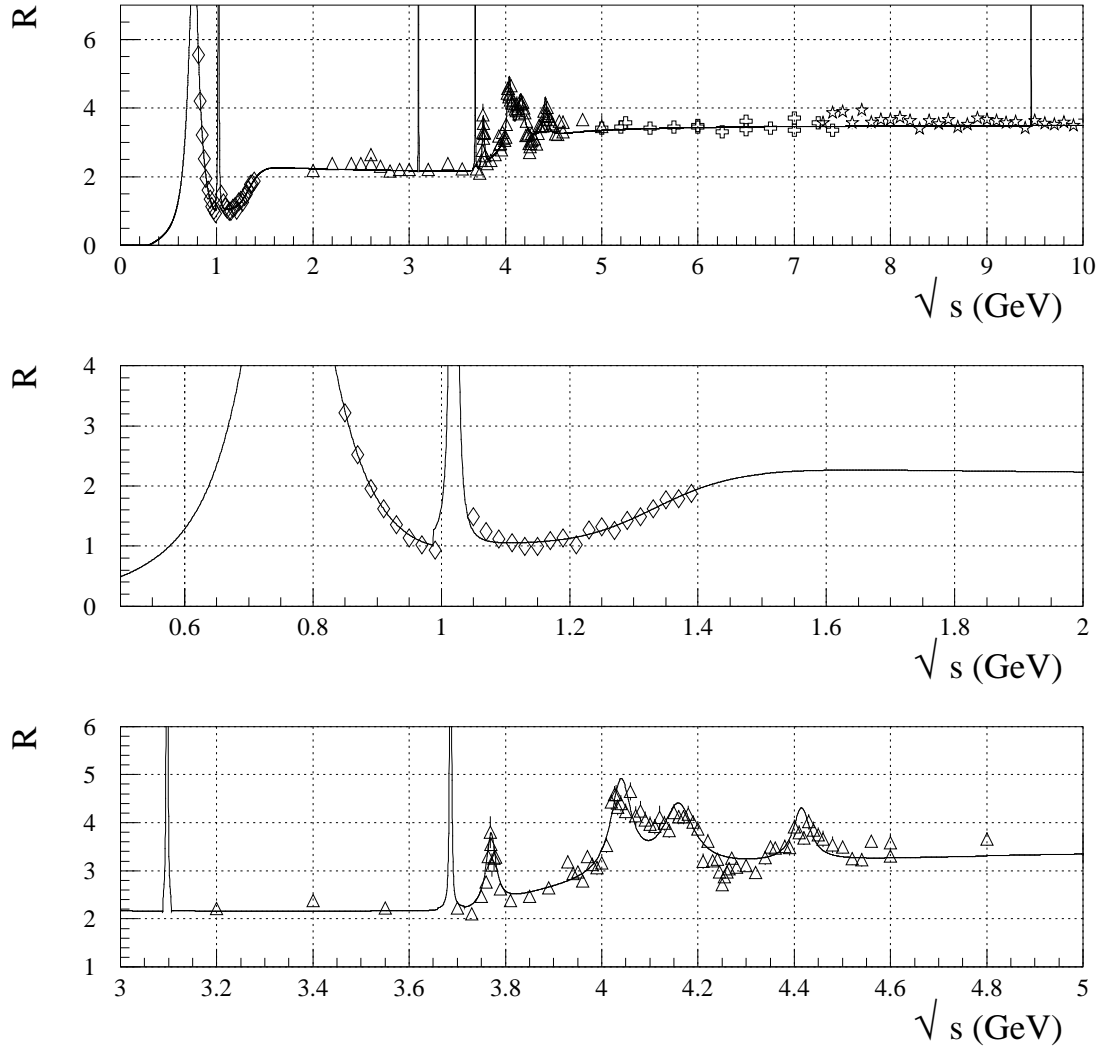,width=16cm}
  \end{center}
  \caption{Parametrization of the R-ratio, and comparison to various
    data sets. The upper figure illustrates the range betweeen 0 and
    10 GeV. The data sets are from ND (below 1.5 GeV), BES (from 2 to
    5 GeV), Crystal Ball (from 5 to 7.4 GeV) and MD-1 (between 7.4 and
    10 GeV) \cite{Rdata}. The middle figure details the $\phi$ region,
    and the lower one shows the $\Psi$ resonance region.}
\label{rparam}
\end{figure}

The resulting parametrization of the R-ratio is illustrated in Figure
\ref{rparam}. 

\subsection{Monte-Carlo sampling}

The R function being very accidented, special attention has to be
paid to its Monte-Carlo sampling. Exploration of all narrow peaks is
performed using iterative dividing of the integration range. In a
first step, the full range is divided in a number of sub-intervals,
the borders of which are set on the (a priori known) peaks of the
resonances. The interval contributing most to the integral is then
divided into two equal sub-intervals. This process is iterated and
fine granularity is obtained by requiring a large number of
intervals. After completion of the initialization phase, drawing
intervals at random (in proportion to their contribution to the total
integral) then ensures correct population of all the peaks. The
program implementing this algorithm and used for our purpose,
{\tt VESKO}, was first described in \cite{jadach}. It was modified to
be able to take into account the positions of a priori known peaks in
the distributions, and to handle several distributions simultaneously.

\subsection{Caveats}

The above procedure is valid only if the final state quark pair is
genuinely coupled to a virtual photon. Therefore, this method should
not be applied to the $e^{+}e^{-} \rightarrow e^{+}e^{-}\mathrm{q\bar{q}}$
process, unless a kinematic region is selected where the multi-peripheral
(mostly $\gamma$-$\gamma$) diagrams have negligible contribution and
interference with the other diagrams.

Other complications appear for final states with four quarks. In
principle, one $\bar{R}$ factor should be present for every pair of
identical quarks, but there are cases where care should be taken. For
example, in the $\mathrm{u\bar{d}d\bar{u}}$ final state and if all
quarks have same colour, one should evaluate whether charged current
diagrams ($ie$, WW production) or neutral current diagrams (Z/$\gamma^*$
Z/$\gamma^*$) dominate the total amplitude, depending on momentum
configurations. Similarly, in final states with four identical quarks
(of same colour), the ``pairing'' which dominates the total amplitude 
has to be determined, and the $\bar{R}$ factors  should be evaluated
at the masses of the connected quark pairs. 

On the contrary, the $e^{+}e^{-} \rightarrow \mathrm{q\bar{q}}$
process is unambiguous. 

\section{Hadronization}
\label{hadron}

Having decided on the $\mathrm{q\bar{q}}$ mass spectrum in the resonance
region, it remains to be determined how every individual event should behave
in the detector. As a function of the mass of the quark pair, it has
to be decided whether we it can be treated as a resonance of known
decay properties, and if not, if the string or cluster
fragmentation algorithms available in {\tt PYTHIA} will work. In the
first case, the quark pair is replaced by the corresponding resonance
in the {\tt PYTHIA} event record, and it is decayed by a subsequent
call to the hadronization routine. In the second case, it is assumed
that above a mass of 2 GeV, string hadronization is convenient, except
in the $\mathrm{c\bar{c}}$ and $\mathrm{b\bar{b}}$ threshold regions
where cluster hadronization correctly takes over.

Since the lowest mass region is dominated by the $\rho$, we are left
with a mass window between $\sim 0.9$ and 2 GeV, open to u-, d-, and s-quarks, 
where new input is needed. This input can be taken from experiment,
since the first hadronic final states in $\mathrm{e^{+}e^{-}}$ collisions
have been extensively studied in the seventies and eighties. Data on the 
following exclusive non-resonant cross-sections \cite{hadrons} were compiled
for this purpose:  
\begin{itemize}
  \item{$ \mathrm{e^{+}e^{-}} \rightarrow 3\pi \,\, (\pi^{+} \pi^{-} \pi^{0}) $,}
  \item{$ \mathrm{e^{+}e^{-}} \rightarrow 4\pi \,\, (2\pi^{+} 2\pi^{-}, \pi^{+} \pi^{-} 2\pi^{0}) $,}
  \item{$ \mathrm{e^{+}e^{-}} \rightarrow 5\pi \,\, (2\pi^{+} 2\pi^{-} \pi^{0}) $,}
  \item{$ \mathrm{e^{+}e^{-}} \rightarrow 6\pi \,\, (2\pi^{+} 2\pi^{-} 2\pi^{0}, 3\pi^{+} 3\pi^{-}) $,}
  \item{$ \mathrm{e^{+}e^{-}} \rightarrow 2K \,\, (K^{+} K^{-}, K_{S} K_{L}) $,}
  \item{$ \mathrm{e^{+}e^{-}} \rightarrow K_{S} K^{+} \pi^{-} $ + CC ,}
  \item{$ \mathrm{e^{+}e^{-}} \rightarrow K^{+} K^{-} \pi^{+} \pi^{-} $.}
\end{itemize}

\noindent The $K_{L} K^{+} \pi^{-}$ final state has never been measured
due to experimental difficulties related to $K_{L}$ detection, but was
assumed to be equal to the $K_{S} K^{+} \pi^{-}$ cross-section.

\begin{figure}
  \caption{The non-resonant ($\rho$ and $\phi$ subtracted) branching
           fractions of the $\gamma^*$ as a function of its mass, from
           measurements of $\mathrm{e^{+}e^{-}}\rightarrow$ exclusive
           final states \cite{hadrons}.} 
  \begin{center}
    \epsfig{figure=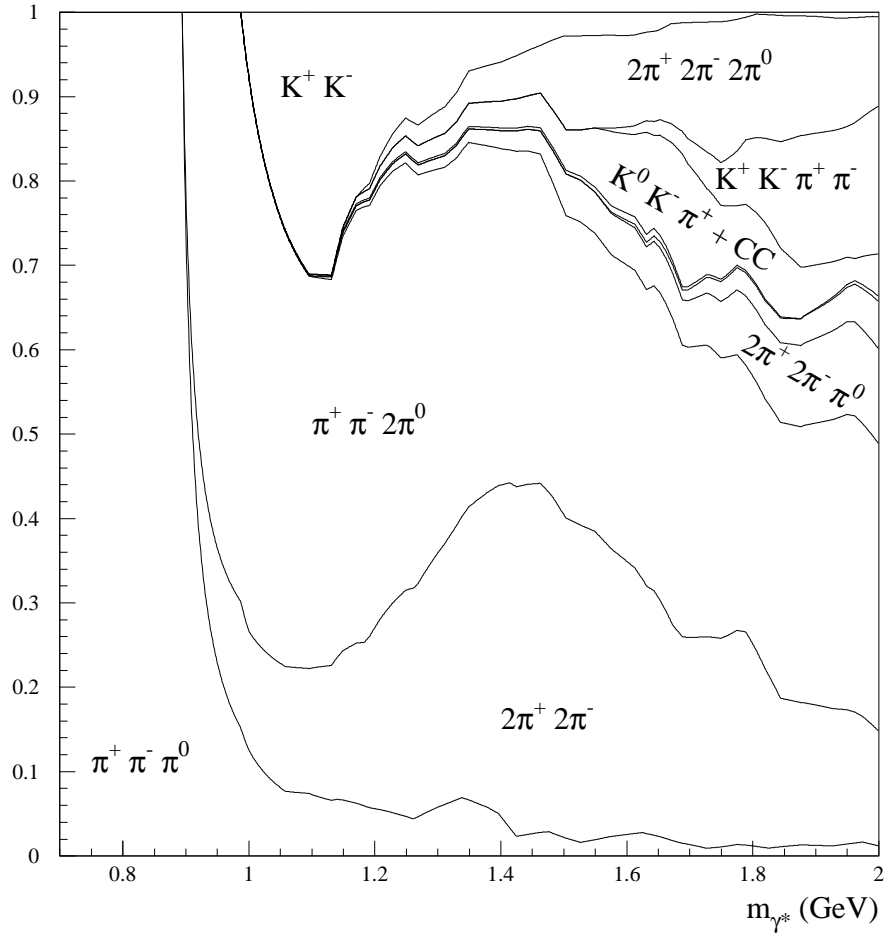,width=12cm}
  \end{center}
\label{BR}
\end{figure}

These cross-sections define the ``branching fractions'' of 
the non-resonant $\gamma^{*}$, which are displayed in figure \ref{BR}. In
practice, when a $\mathrm{q\bar{q}}$ pair is found to be neither a 
resonance, nor heavy enough to be administrated by string fragmentation, 
one of the above final states is drawn at random according to the
measured cross-sections at the current $\mathrm{q\bar{q}}$ mass. The
momenta of the outgoing hadrons are then generated uniformly in the
available phase space using the CERN routine {\tt GENBOD}
\cite{fjames} (it has currently not been attempted to reproduce the
inner kinematic structure of these multiparticle final states. For
example, it is ignored that a significant part of the
$2\pi^{+}2\pi^{-}$ final state is genuinely $\rho\rho$
decays). Finally, the quark pair is replaced by its daughter hadrons
in the {\tt PYTHIA} event record.

\section{Results for specific processes}
\label{someres}

\begin{figure}
  \caption{Differential radiative cross-sections expected at the
    $\Psi$(3S) resonance (left) and at the $\Upsilon$(4S) resonance
    (right). No selections are applied.}
  \begin{center}
    \begin{tabular}{cc}
      \epsfig{figure=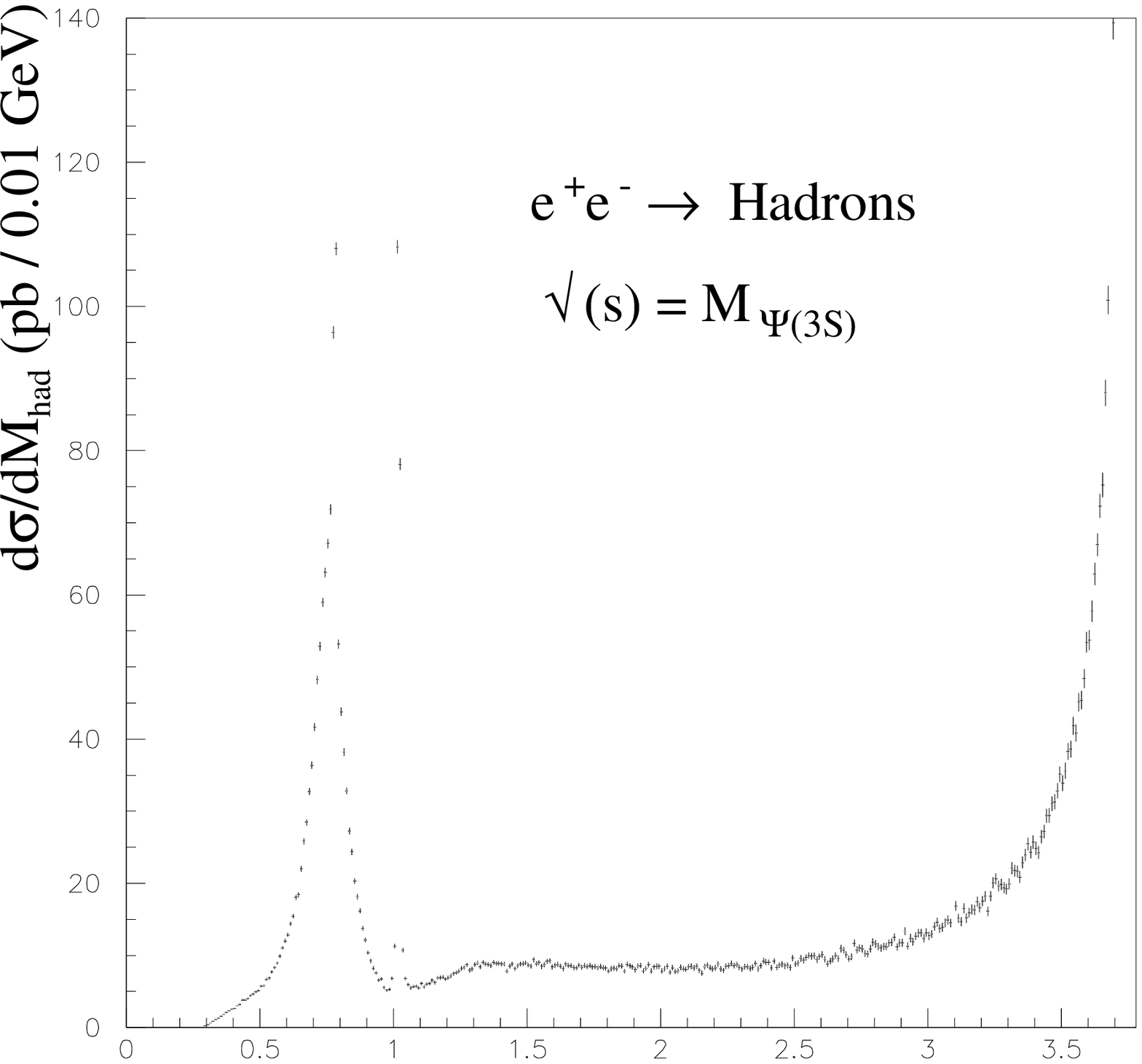,width=7.8cm} &
      \epsfig{figure=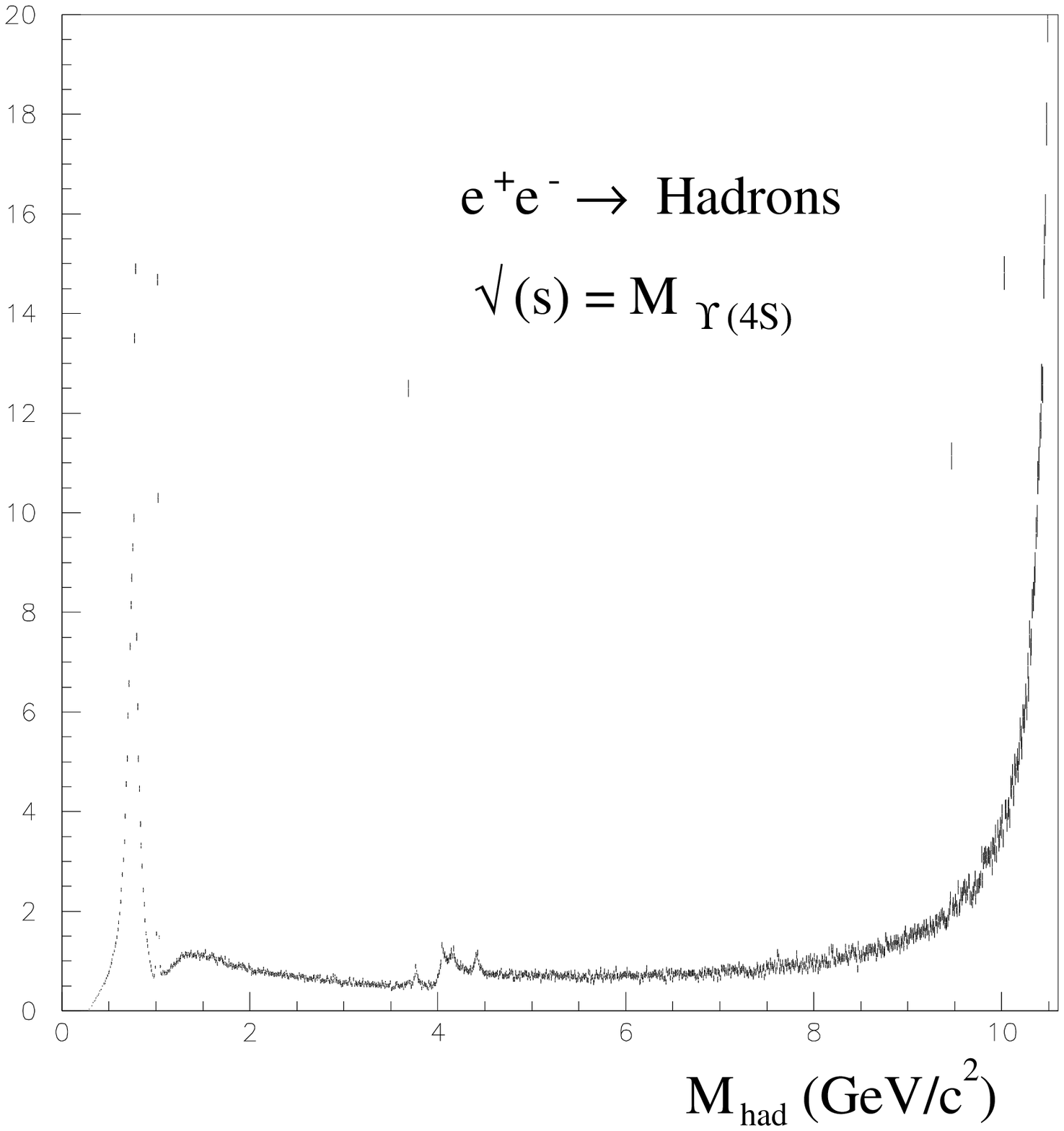,width=7.3cm}
    \end{tabular}
  \end{center}
\label{spectrum}
\end{figure}

The present algorithm has been implemented in the 2-fermion
generator {\tt KK2f} \cite{kk2f}. Running this generator in the
configuration of current $e^{+}e^{-}$ colliders allows to estimate the
radiative hadronic cross-sections expected at these machines. Table
\ref{csec} summarizes results for the existing $\Phi$- and
B-factories, and for a hypothetical Charm factory such as the
Cleo-c program at CESR. Given the expected luminosities at these machines, 
${\mathcal O}(10^6-10^7)$ events are expected in regions where the knowledge of
the hadronic cross-section is still limited\,\footnote{These numbers
  are given with initial state radiation computed to third order (see
  \cite{kk2f}), and final state radiation switched off. A discussion
  on the level of accuracy of this approximation is beyond the scope
  of this work, but is expected to be 1-2\%.}.
Figure \ref{spectrum} illustrates the differential hadronic
cross-sections expected when operating at D- and B-meson threshold.

\begin{table}[!ht]
  \caption{Expected $e^{+}e^{-} \rightarrow \mathrm{q\bar{q}}$ event
    rates (in nanobarn) for various colliders. Cross-sections
      are given for three ranges of the effective center-of-mass
      energy (for the hypothetical Charm factory operating on the
      $\Psi(3S)$, the last number corresponds to the range 2-3
      GeV). No selections are applied.}
  \begin{center}
    \begin{tabular}{ccccc}
      Collider      &  $\sqrt{s}$ (GeV)  & 0-0.95 GeV  &  1.05-2 GeV &  2-5 GeV \\
      \hline \\
      DA$\Phi$NE    &  1.02 & 42.38  &      & \\
      Charm factory &  3.77 &  1.31  & 0.77 & $0.95^{\star}$ \\
      KEKB,PEP-II   & 10.58 &  0.18  & 0.09 & 0.25 \\
    \end{tabular}
  \end{center}
\label{csec}
\end{table}

In the context of four-fermion processes, events with a light
quark-pair will show up as thin, collimated jet (most often a
$\pi^{+}\pi^{-}$ pair from a $\rho$ decay) recoiling against an
energetic lepton- or jet-pair. Table \ref{csec4f} summarizes some
results for the $\mu^{+}\mu^{-}\mathrm{q\bar{q}}$ and
$\mathrm{q\bar{q}q\bar{q}}$ processes. Assuming 500 $\mathrm{fb^{-1}}$
recorded per experiment, the number of expected events with one quark
pair of mass smaller than 2 GeV is ${\mathcal O}(10^1-10^2)$,
depending on the process under consideration. This confirms that such
events can represent a substantial background to new particle
searches, and need to be accounted for in the simulation of standard
model processes. 

\begin{table}[!ht]
  \caption{Expected $e^{+}e^{-} \rightarrow
    \mu^{+}\mu^{-}\mathrm{q\bar{q}}$ and  $\mathrm{q\bar{q}q\bar{q}}$
    event rates (in picobarn) at LEP2 energies. Quark-pair masses are
    restricted above 2 GeV (case 1), and released down to threshold (case 2).}
  \begin{center}
    \begin{tabular}{cccc}
      Final state         &  $\sqrt{s}$ (GeV)  & Case 1 & Case 2 \\
      \hline \\
      $\mu^{+}\mu^{-}\mathrm{q\bar{q}}$ & 200 &  $0.33 $ & $0.36 $ \\
      $\mathrm{q\bar{q}q\bar{q}}$       & 200 & $8.72 $ & $9.05 $ \\
    \end{tabular}
  \end{center}
  \label{csec4f}
\end{table}

\section{Program availability}
\label{example}

The routines implementing the methods described above are obtainable
from \cite{rescode}, together with installation notes and commented example
applications.


\section{Conclusions}

A complete treatment of low-mass quark pair production and
hadronization is described, which is relevant whenever the pair is
produced via a virtual photon. This allows to release traditional cuts on
quark pair invariant masses, and to study low mass hadronic systems
specifically. The large statistics available at existing $e^{+}e^{-}$
machines will allow to compare these predictions with experiment, with
applications both in two- and four-fermion processes.



\end{document}